\documentclass[aps,prl,reprint,groupedaddress]{revtex4-1}
\usepackage{graphicx} 
\usepackage{dcolumn} 
\usepackage{bm}
\usepackage{hyperref}

\begin{document}

\title{Antiproton production in heavy-ion collisions at subthreshold energies}
\author{Zhao-Qing Feng}
\email{Corresponding author: fengzhq@scut.edu.cn}

\affiliation{School of Physics and Optoelectronics, South China University of Technology, Guangzhou 510640, China}

\date{\today}

\begin{abstract}
Within the framework of the Lanzhou quantum molecular dynamics model, the deep subthreshold antiproton production in heavy-ion collisions has been investigated thoroughly. The elastic scattering, annihilation and charge exchange reactions associated with antiproton channels are implemented in the model. The attractive antiproton potential extracted from the G-parity transformation of the nucleon selfenergies enhance the antiproton yields to some extent. The calculated invariant spectra are consistent with the available experimental data. The primordial antiproton yields increase with the mass number of colliding system. However, the annihilation reactions reduce the antiproton production and enable independent to colliding partners. Antiflow phenomena of antiprotons correlated to the mean-field potential and annihilation mechanism is found in comparison with proton flows.

\begin{description}
\item[PACS number(s)]
25.75.Dw, 25.43.+t, 25.75.Ld
\end{description}
\end{abstract}

\maketitle

\section{I. Introduction}

Heavy-ion collisions in terrestrial laboratory provide a unique possibility to create the dense hadronic matter, which are available for investigating the in-medium properties of hadrons and nuclear equation of state \cite{Ca99, Fu06, Li08, To10, Ha12}. Particle production at energies below the threshold for its creation in nucleon-nucleon collisions can be probe to study the high-density hadronic matter properties, i.e., the chiral symmetry restoration, phase-transition from quark-gluon plasma to hadrons, hadron-nucleon interaction, nuclear equation of state etc \cite{Gi95, Fr07, Fe13, Lu17,Fe18}. A number of experiments for the subthreshold production of pions, kaons, antikaons and antiprotons in heavy-ion collisions were performed and precise spectra have been measured \cite{Ho96,La99,Ca89}. The in-medium properties associated with the secondary reactions, propagation and quasiparticle concept in matter were investigated thoroughly. The subthreshold antiproton production is more complicated because of the annihilation process. The first evidence of antiproton production was dated in 1955 at Berkeley in collisions of protons on copper at the energy of 6.2 GeV by Chamberlain, Segr\`{e}, Wiegand and Ypsilantis \cite{Ch55}. The first experiments at BEVALAC and JINR in the 1980s were performed for the subthreshold antiproton production in heavy-ion collisions and followed by precise measurements at KEK and GSI. Recently, the antiproton pair correlation was investigated by the STAR collaboration in relativistic heavy-ion collisions \cite{Star}. The secondary beams with antiprotons were produced at many laboratories such as CERN, BNL, KEK \cite{Ch57,Ag60,Le80,Ea99} and being constructed at PANDA (Antiproton Annihilation at Darmstadt, Germany) for the hypernuclear physics, charmonium physics, hadron spectroscopy etc. The antiproton physics is also being planned in the future experiments at the high-intensity heavy-ion accelerator facility (HIAF) in Huizhou, China \cite{Ya13}.

The antiproton production in heavy-ion collisions or proton induced reactions at deep subthreshold energies is related to the antiproton-nucleon interaction and also coupled to a number of reaction channels, i.e., the meson-baryon and baryon-baryon collisions, annihilation channels, charge-exchange reaction, elastic and inelastic collisions associated with antiprotons. There has been several approaches for describing the antiproton production, e.g., the fireball model \cite{Ko89}, the first-chance nucleon-nucleon collision model \cite{Sh90}, the quasicoherent multiparticle collision model \cite{Da90} and the microscopic transport approaches \cite{Ba91,Hu92,Li94}. These models can explain the antiproton spectra in proton-nucleus and nucleus-nucleus collisions to some extent. Self-consistent treatment of all possible channels contributing the antiproton production is still necessary, in particular the secondary reactions with annihilation products.

In this work, the microscopic mechanism of antiproton production in heavy-ion collisions at subthreshold energies is to be investigated with the Lanzhou quantum molecular dynamics (LQMD) transport model. The article is organized as follows. In section II we give a brief description of the extension for antiproton production. The calculated results and discussion are presented in section III. Summary and perspective on the antiproton physics are outlined in section IV.

\section{II. Brief description of the LQMD transport model}

In the Lanzhou quantum molecular dynamics (LQMD), the dynamics of resonances with the mass below 2 GeV, hyperons ($\Lambda$, $\Sigma$, $\Xi$) and mesons ($\pi$, $\eta$, $K$, $\overline{K}$, $\rho$, $\omega$) is associated with the reaction channels and mean-field potential, which is coupled each other in the hadron-hadron collisions, antibaryon-baryon annihilations, decays of resonances \cite{Fe11,Fe12}. The temporal evolutions of all particles are described by Hamilton's equations of motion under the self-consistently generated mean-field potentials. In this work, the antiproton production in nucleon-nucleon collisions is implemented in the model. The antiproton-nucleon potential is evaluated from the dispersion relation as
\begin{eqnarray}
V_{opt}(\textbf{p},\rho)=\omega_{\overline{B}}(\textbf{p}_{i},\rho_{i}) - \sqrt{\textbf{p}^{2}+m^{2}}          \\
\omega_{\overline{B}}(\textbf{p}_{i},\rho_{i})=\sqrt{(m_{\overline{B}}+\Sigma_{S}^{\overline{B}})^{2}+\textbf{p}_{i}^{2}} + \Sigma_{V}^{\overline{B}}
\end{eqnarray}
with $\Sigma_{S}^{\overline{B}}=\Sigma_{S}^{B}$ and $\Sigma_{V}^{\overline{B}}=-\Sigma_{V}^{B}$.
The nuclear scalar $\Sigma_{S}^{N}$ and vector $\Sigma_{V}^{N}$ self-energies are computed from the well-known relativistic mean-field model with the NL3 parameter. The relativistic self-energies are used for the construction of hyperon and antibaryon potentials only. A factor $\xi$ is introduced in order to fit the phenomenological optical potential as $\Sigma_{S}^{\overline{N}}=\xi\Sigma_{S}^{N}$ and $\Sigma_{V}^{\overline{N}}=-\xi\Sigma_{V}^{N}$ with $\xi$=0.25, which leads to the strength of $V_{\overline{N}}=-164$ MeV at the normal nuclear density $\rho_{0}$=0.16 fm$^{-3}$. The effective mass $m^{\ast}_{\overline{p}}=\omega_{\overline{B}}(\textbf{p}=0,\rho=\rho_{0})$ is used to evaluate the threshold energy for antiproton production, e.g., the threshold energy in the nucleon-nucleon collisions $\sqrt{s_{th}}=m^{\ast}_{\overline{p}} + 3m_{N}$ and $m_{N}$ being the mass of nucleon.

Production and decay of resonances in meson-baryon and baryon-baryon collisions have been implemented into in the LQMD model \cite{Fe12}, in which the strangeness and vector mesons are created via direct process. The antiproton production is related to the pion-baryon and nucleon-baryon channels at the subthreshold energy (E$_{th}$=5.62 GeV) as
\begin{equation}
\pi B \rightarrow Np\overline{p}, BB \rightarrow NN p\overline{p}.
\end{equation}
Here the cross sections in the pion-baryon and nucleon-baryon channels are evaluated with the same form of \cite{Si98}
\begin{equation}
\sigma_{\pi (B)B \rightarrow \overline{p}X}(\sqrt{s}) = a\left(\frac{s}{s_{0}}-1\right)^{b}\left(\frac{s_{0}}{s}\right)^{c}
\end{equation}
with parameters of $a=1$ mb, $b=2.31$, $c=2.3$ and $a=0.12$ mb, $b=3.5$, $c=2.7$ for the pion and nucleon induced reactions, respectively. Isotropic distribution for the antiproton production is included.

The annihilation reactions in antibaryon-baryon collisions are described by a statistical model with SU(3) symmetry of pseudoscalar and vector mesons \cite{Go92}, which considers possible combinations with the final state from two to six mesons \cite{La12}. Besides the annihilation channels, charge-exchange reaction (CEX), elastic (EL) and inelastic scattering with antibaryons are also implemented in the model as follows \cite{Fe14}.
\begin{eqnarray}
&& \overline{B}B \rightarrow \texttt{annihilation}(\pi,\eta,\rho,\omega,K,\overline{K},\eta\prime,K^{\ast},\overline{K}^{\ast},\phi),
\nonumber  \\
&&  \overline{B}B \rightarrow \overline{B}B (\texttt{CEX, EL}),   \overline{N}N \leftrightarrow \overline{N}\Delta(\overline{\Delta}N), \overline{B}B \rightarrow \overline{Y}Y.
\nonumber  \\
\end{eqnarray}
Here the B strands for nucleon and $\Delta$(1232), Y($\Lambda$, $\Sigma$, $\Xi$), K(K$^{0}$, K$^{+}$) and $\overline{K}$($\overline{K^{0}}$, K$^{-}$). The overline of B (Y) means its antiparticle. The cross sections of these channels are based on the parametrization or extrapolation from available experimental data.

\section{III. Results and discussion}

Particles produced in heavy-ion collisions can manifest the properties of high-density hadronic matter formed at the compression stage. Shown in Fig. 1 is the time evolution of mesons ($\pi^{+}$, $\eta$, K$^{+}$ and K$^{-}$), hyperons ($\Lambda$, $\Sigma^{-}$ and $\Xi^{-}$), primordial and annihilation antiprotons in central $^{28}$Si+$^{28}$Si collisions at 3 GeV/nucleon. It is obvious that most of particles equilibrate at the collision stage (after 20 fm/c) except for K$^{-}$ and annihilation antiprotons. Strangeness exchange process $K^{-}N\rightarrow \pi Y$ and annihilation reactions reduces the yields of K$^{-}$ and antiprotons, respectively. Pions are main products in the antiproton annihilation process. Therefore, the pion induced reactions contribute the strange particle production. The primordial antiprotons with the multiplicity of $10^{-5}$ are reduced to be about $4\times10^{-7}$. Roughly, only about $4\%$ of the primordial antiprotons can be emitted from the dense hadronic matter, which is consistent with the calculations by relativistic Vlasov-Uehling-Uhlenbeck (RVUU) model \cite{Li94}. The incident energy dependence of the primordial and annihilation antiprotons is calculated as shown in Fig. 2. The escaping probability from the annihilation varies from $1.3\%$ at 2 GeV/nucleon to $7.5\%$ at 4 GeV/nucleon. The collision time is shorten with increasing the beam energy, which is available for the antiproton escape from the hadronic matter.

\begin{figure*}
\includegraphics[width=16 cm]{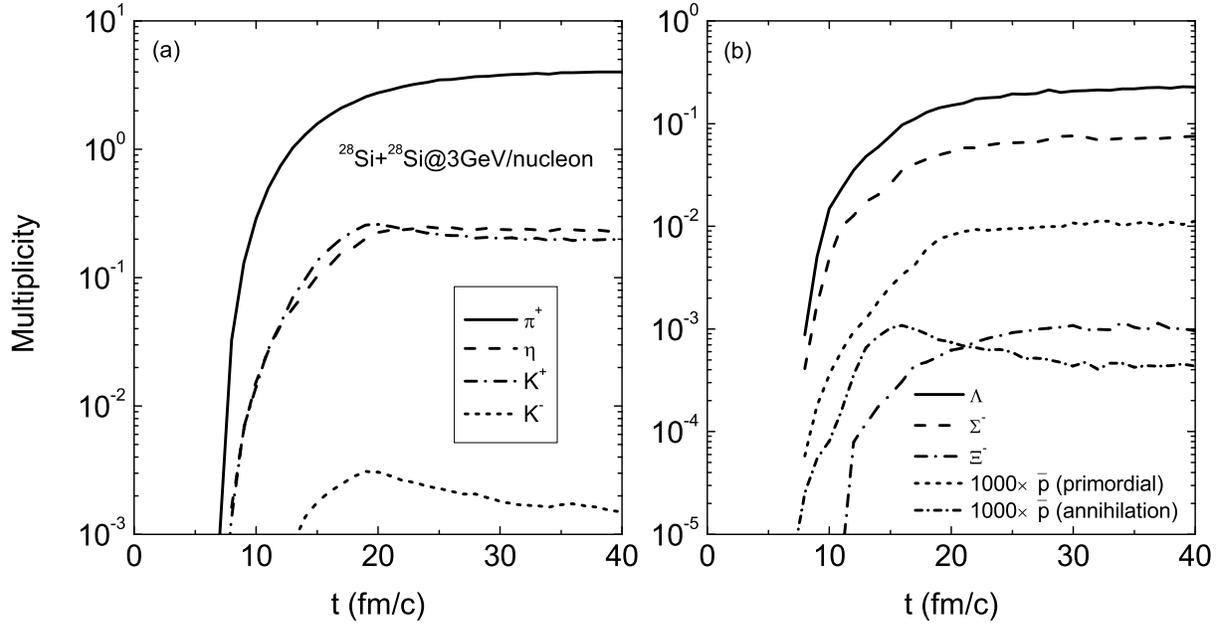}
\caption{Temporal evolution of the mesons, hyperons, primordial and annihilation antiprotons produced in $^{28}$Si+$^{28}$Si collisions at the incident energy of 3 GeV/nucleon.}
\end{figure*}

\begin{figure*}
\includegraphics[width=16 cm]{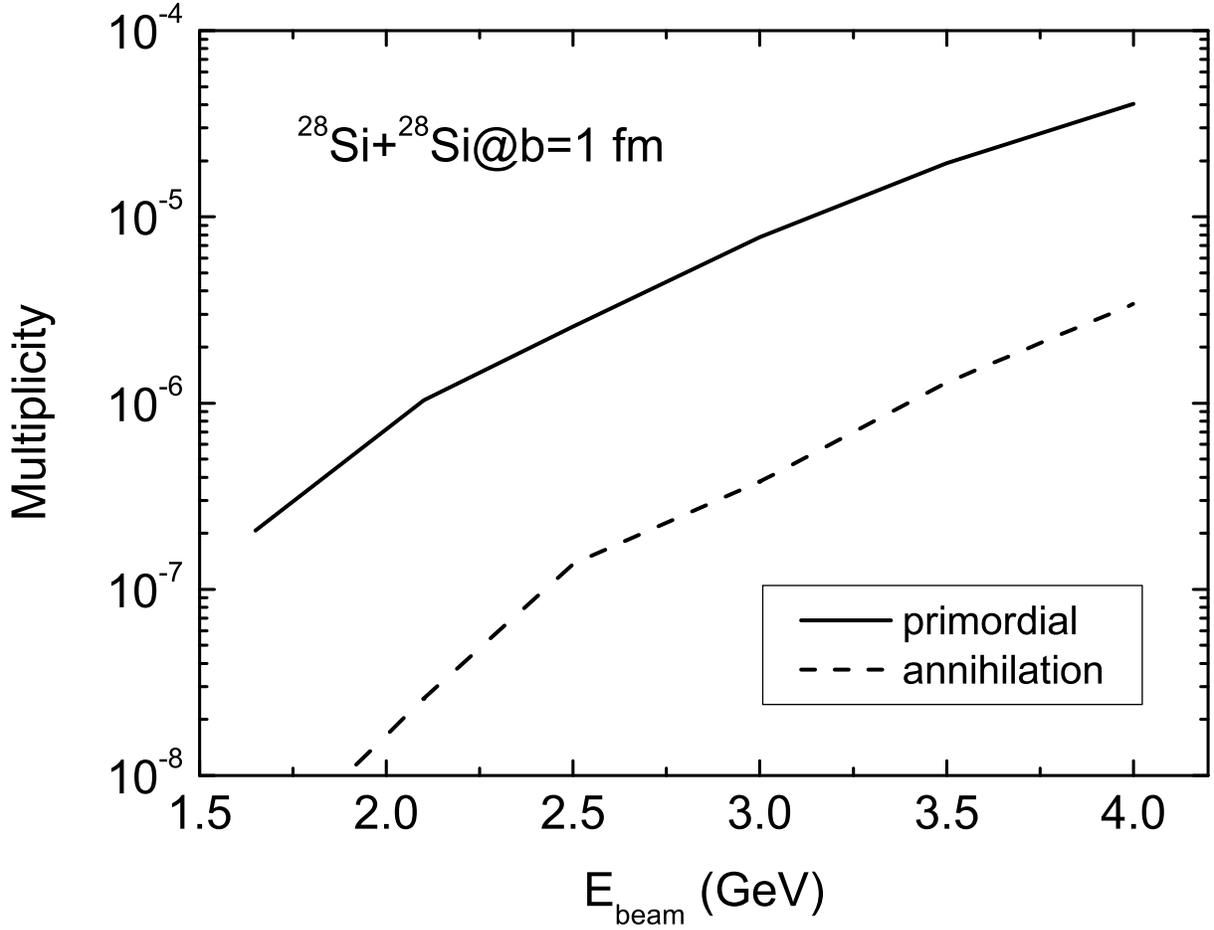}
\caption{The primordial and annihilation antiprotons as a function of the incident energy in central $^{28}$Si+$^{28}$Si collisions.}
\end{figure*}

The hadronic matter formed in heavy-ion collisions has been manifested to be related to the reaction system. Heavy nuclei induced reactions enhance the baryon density in the colliding region, which enable the subthreshold particle production and more pronounced in-medium effects. Systematic analysis is useful for reducing some uncertainty quantities, such as the in-medium cross section, decay width etc. We calculated the K$^{+}$, K$^{-}$, primordial and annihilation antiprotons at 2 GeV/nucleon in the central collisions of $^{12}$C+$^{12}$C, $^{28}$Si+$^{28}$Si, $^{40}$Ca+$^{40}$Ca, $^{58}$Ni+$^{58}$Ni, $^{112}$Sn+$^{112}$Sn and $^{197}$Au+$^{197}$Au as shown in Fig. 3. It is obvious that the yields of K$^{+}$ and primordial antiprotons increase monotonically with the mass number of colliding system. The participant nucleons are available for particle production. However, a platform appears for K$^{-}$ and annihilation antiprotons because of the secondary collisions. The antiproton-nucleon potential and annihilation effects are analyzed as shown in Fig. 4. The inclusion of the mean-field potential enhances the antiproton production owing to the reduction of threshold energy. The underestimation of the antiproton yields in nucleus-nucleus collisions without the correction on threshold energy was also found in other models. On the other hand, the attractive $\overline{p}$-N potential leads to the reduction of high-momentum antiprotons. The contribution of the annihilation channel reduces a number of antiprotons, in particular in the domain of low momentum. The enhancement of subthreshold antiproton yields in deuteron induced reactions was found in comparison to the ones in proton-nucleus collisions at KEK \cite{Ch93}. The annihilation mechanism with high-intensity antiproton beams is to be investigated at PANDA in the near future experiments.

\begin{figure*}
\includegraphics[width=16 cm]{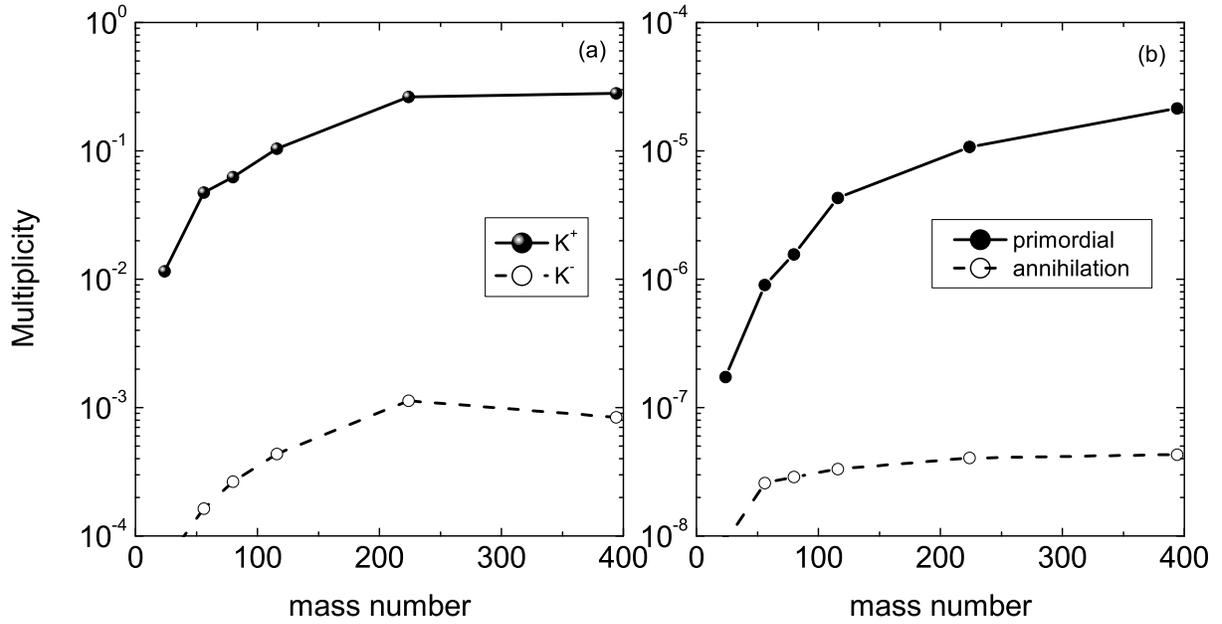}
\caption{Mass dependence of K$^{+}$, K$^{-}$, primordial and annihilation antiprotons at the incident energy of 2 GeV/nucleon.}
\end{figure*}

\begin{figure*}
\includegraphics[width=16 cm]{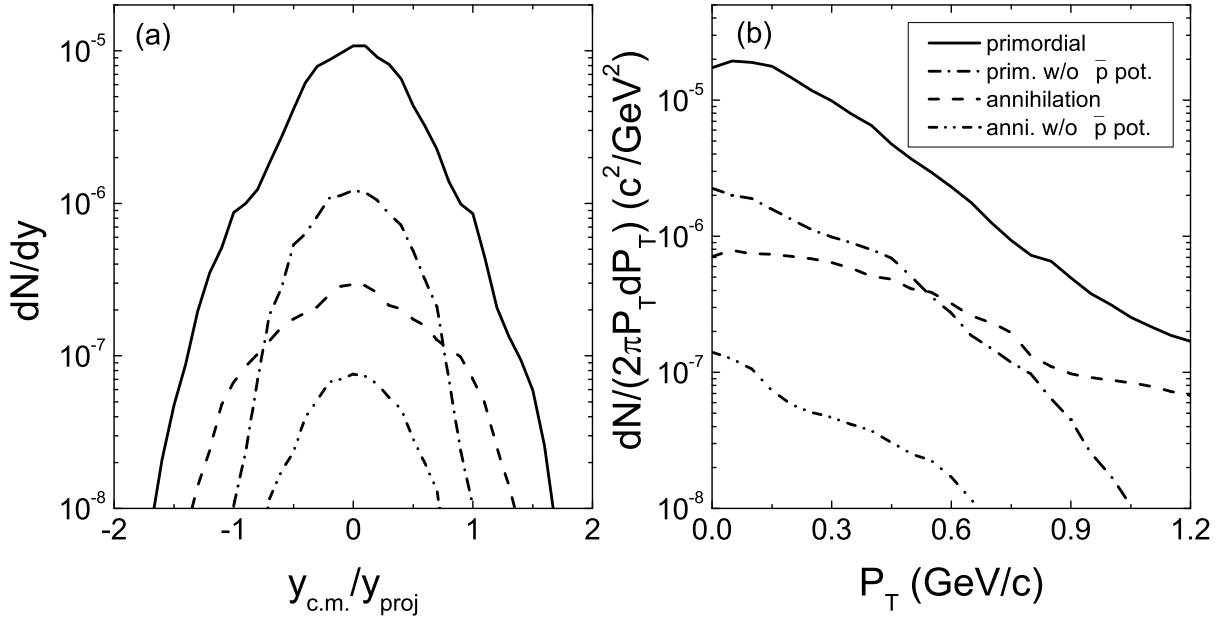}
\caption{Rapidity and transverse momentum spectra primordial and annihilation antiprotons in central $^{40}$Ca+$^{40}$Ca collisions at the energy of 3 GeV/nucleon.}
\end{figure*}

The kinetic energy spectra of invariant cross section can be probes of the properties of hadronic matter, i.e., the local temperature of particle emission, particle optical potential, nuclear equation of state etc. Shown in Fig. 5 is the inclusive spectra of antiprotons produced in collisions of $^{28}$Si+$^{28}$Si and $^{58}$Ni+$^{58}$Ni at the incident energies of 2 GeV/nucleon and 1.85 GeV/nucleon compared with the available experimental data, respectively \cite{Sh89,Sc94}. Calculated results are consistent with the data when the annihilation channel is included. The primordial antiprotons are emitted at the early stage in nucleus-nucleus collisions. The annihilation widely reduces the antiproton production in the whole energy range and leads to the creation of pions in the dense matter. The collisions of the pions with the surrounding nucleons might produce antiprotons again. The multiple process increases the antiproton production to some extent. The antiproton production in heavy-ion collisions is related to the collision centrality \cite{Be95}. The central collisions are available the antiproton annihilation. Consequently, the antiproton yields fail to follow the linear dependence with the collision centrality.

\begin{figure*}
\includegraphics[width=16 cm]{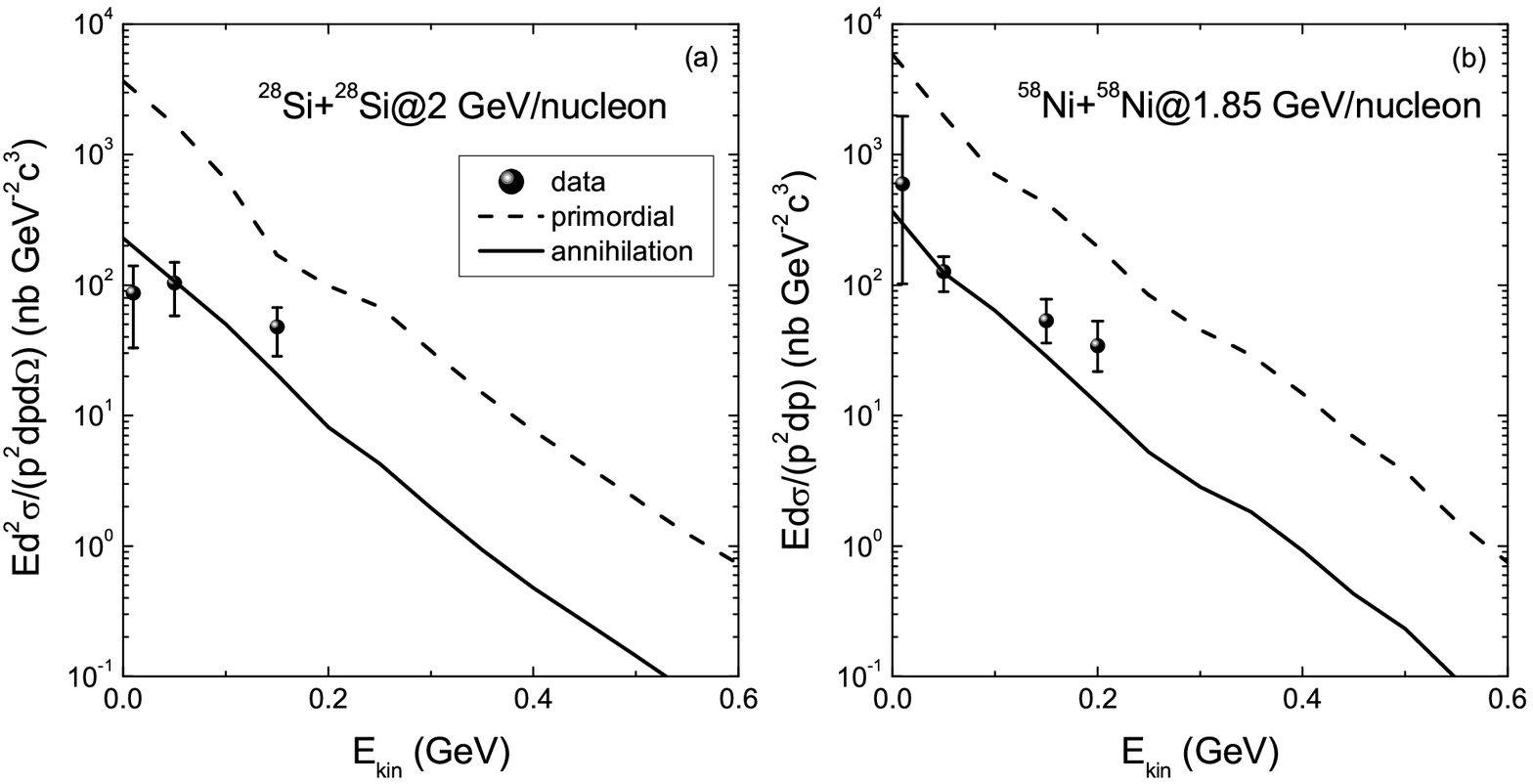}
\caption{Invariant spectra of antiprotons produced in collisions of $^{28}$Si+$^{28}$Si and $^{58}$Ni+$^{58}$Ni at the incident energies of 2 GeV/nucleon and 1.85 GeV/nucleon compared with the available experimental data, respectively \cite{Sh89,Sc94}.}
\end{figure*}

Collective flows in heavy-ion collisions provide the azimuthal correlations of emitted particles, which have been used for extracting the properites of high-density baryonic matter. The flow information can be exhibited from the Fourier expansion in phase space, i.e., expressed as $\frac{dN}{d\phi}(y,p_{t})=N_{0}(1+2V_{1}(y,p_{t})\cos(\phi)+2V_{2}(y,p_{t})\cos(2\phi)+\ldots)$ , where the $p_{t}=\sqrt{p_{x}^{2}+p_{y}^{2}}$ and $y$ are the transverse momentum and the longitudinal rapidity along the beam direction, respectively. The directed (transverse) flow is defined as the first coefficient and expressed as $V_{1}=\langle p_{x}/p_{t} \rangle$, which provides the information of the azimuthal anisotropy of the transverse emission. The elliptic flow $V_{2}=\langle (p_{x}/p_{t})^{2}-(p_{y}/p_{t})^{2} \rangle$ gives the competition between the in-plane ($V_{2}>$0) and out-of-plane ($V_{2}<$0, squeeze out) emissions. The brackets indicate averaging over all events in accordance with a specific class such as rapidity or transverse momentum cut. The transverse flows of nucleons, light clusters, pions and strange particles in heavy-ion collisions have been investigated for probing the high-density symmetry energy, in-medium NN cross section, optical potentials of particles in nuclear matter, particle emission etc \cite{Fe13,Fe12}. To investigate the antiproton azimuthal distribution in the reaction plane, we calculated the spectra of $\pi^{+}$, K$^{+}$ and antiprotons produced in semicentral collisions of $^{58}$Ni+$^{58}$Ni (b=4 fm) at the incident energy of 3 GeV/nucleon as shown in Fig. 6. It is noticed that the directed flows of $\pi^{+}$, K$^{+}$ and antiprotons are anti-correlated in comparison to protons. The antiflow effect is cause from the strongly attractive interaction between antiprotons and nucleons. It has been known that the reabsorption process of $\pi^{+}$ by participant nucleons (shadowing effect) contributes the antiflow phenomena. The anti-correlation of directed flow for antiproton production in heavy-ion collisions was measured at AGS from the transverse momentum spectra \cite{Ba00}.

\begin{figure*}
\includegraphics[width=16 cm]{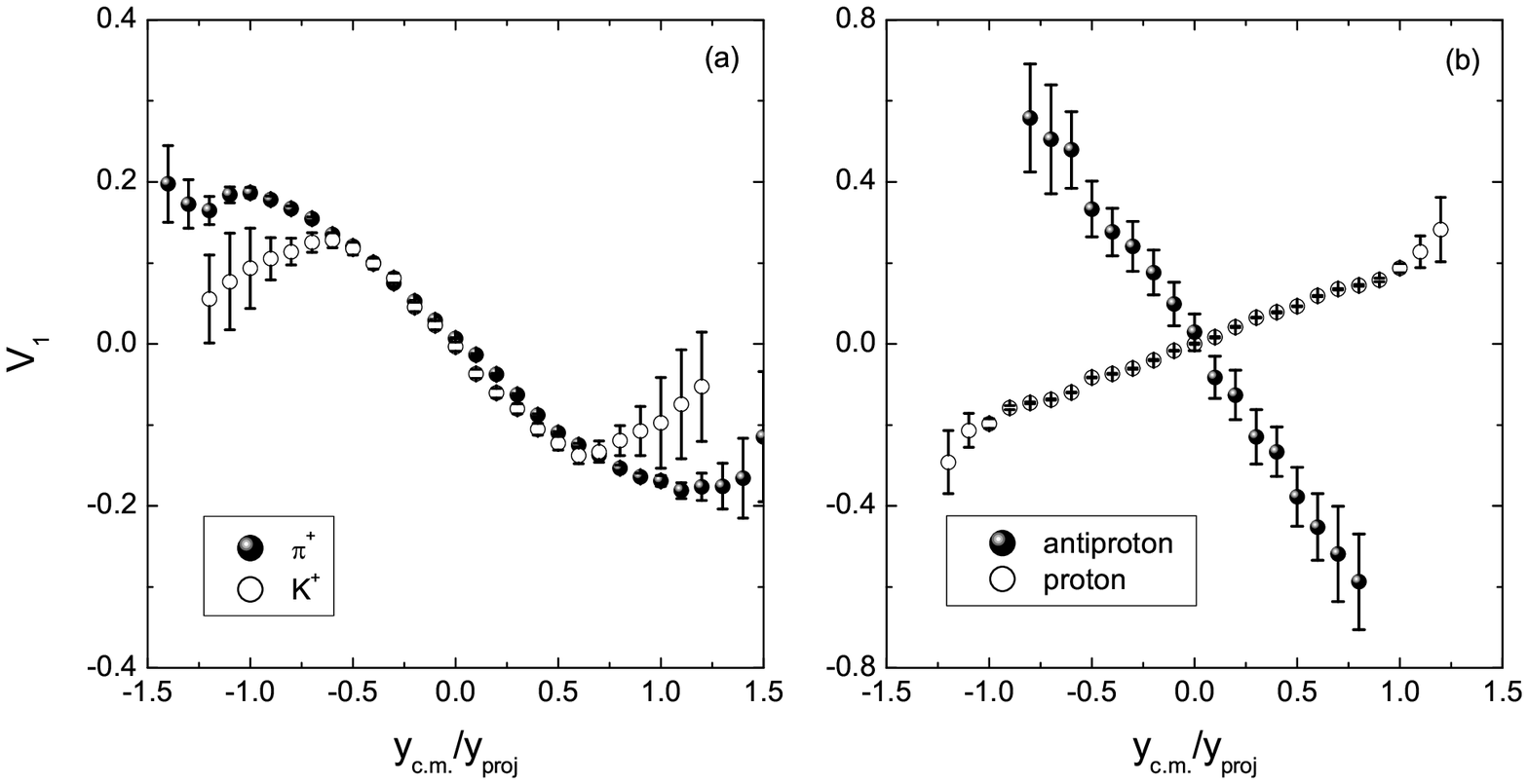}
\caption{Rapidity distributions of direct flows of $\pi^{+}$, K$^{+}$ and antiprotons produced in $^{58}$Ni+$^{58}$Ni at the incident energy of 3 GeV/nucleon.}
\end{figure*}

\section{IV. Conclusions}

In summary, the antiproton dynamics in heavy-ion collisions at deep subthreshold energies has been investigated with the LQMD transport model. The collective effect of antiproton production in heavy-ion collisions is obvious than the ones in proton induced reactions. The influence of the annihilation and $\overline{p}$-N potential on antiproton production is analyzed thoroughly. The inclusion of the $\overline{p}$-N potential enhances the antiproton abundance because of the reduction of threshold energy. The available experimental data of invariant spectra are well reproduced with the model after taking into account the annihilation and optical potential. The pion-nucleon channel slightly enhances the antiproton production. The directed flow of antiprotons is anticorrelated to the proton flow in heavy-ion collisions.

\section{Acknowledgements}

This work was supported by the National Natural Science Foundation of China (Projects No. 11722546 and No. 11675226) and the Talent Program of South China University of Technology.

\end{document}